# Newly synthesized 3D boron-rich chalcogenides $B_{12}X$ (X = S, Se): Theoretical characterization of physical properties for optoelectronic and mechanical applications

M. M. Hossain[1,*], M. A. Ali[1], M. M. Uddin[1], S. H. Naqib[2,*], A. K. M. A. Islam[2,3,*]

[1]Department of Physics, Chittagong University of Engineering and Technology (CUET), Chattogram-4349, Bangladesh
[2]Department of Physics, University of Rajshahi, Rajshahi 6205, Bangladesh
[3]Department of Electrical and Electronic Engineering, International Islamic University Chittagong, Kumira, Chittagong, 4318, Bangladesh

Corresponding authors; [*]email: mukter_phy@cuet.ac.bd; salehnaqib@yahoo.com; azi46@ru.ac.bd

**Abstract**

Boron rich chalcogenides have been predicted to have excellent properties for optical and mechanical applications in recent times. In this regard, we report the electronic, optical and mechanical properties of recently synthesized boron rich chalcogenide compounds, $B_{12}X$ (X = S and Se) using density functional theory for the first time. The effects of exchange and correlation functional on these properties are also investigated. The consistency of the obtained crystal structure with the reported experimental results has been checked in terms of lattice parameters.The considered materials are mechanically stable, brittle and elastically anisotropic. Furthermore, the elastic moduli and hardness parameters are calculated, which show that $B_{12}S$ is likely to be a prominent member of hard materials family compared to $B_{12}Se$. The origin of different in hardness is explained on the basis of density of states near the Fermi level. Reasonably good values of fracture toughness and machinability index for $B_{12}X$ (X= S and Se) are reported. The melting point, $T_m$ for the $B_{12}S$ and $B_{12}Se$ compounds suggests that both solids are stable, at least up to 4208 and 3577 K, respectively. Indirect band gap of $B_{12}S$ (2.27 eV) and $B_{12}Se$ (1.30 eV) are obtained using the HSE06 functional. The energy gap using LDA and GGA are found to be significantly lower. The electrons of $B_{12}Se$ compound show lighter average effective mass compared to that of $B_{12}S$ compound, which signifies higher mobility of charge carriers in $B_{12}Se$. The optical properties such as dielectric function, refractive index, absorption coefficient, reflectivity and loss function are characterized using GGA-PBE and HSE06 method and discussed in detail. These compounds possess bulk optical anisotropy and excellent absorption coefficients in visible light region along with very low static value of reflectivity spectra (range: 7.42-14.0% using both functionals) are noted. Such useful features of the compounds under investigation show promise for applications in optoelectronic and mechanical sectors.

## 1. Introduction

Boron (B)-rich compounds having structural unit of $B_{12}$ closed cage clusters are novel systems attracting significant interest of the research community.[1–6] Many interesting physical properties such as high hardness, useful electronic properties, low mass density, very high melting point, and excellent thermal stability and chemically inertness of B-rich compounds have gained extra attention of the materials science and engineering research community in view of the potential technical applications.[1,6] Among the B-rich compounds, the B-rich solids with icosahedral cluster[1,6,7] structurally derived from α-rhombohedral boron (α-rh boron) offers some industrially potential candidate materials similar to boron carbide, boron sub-oxide and boron sub-phosphide, $REB_{15.5}CN$, $REB_{22}C_2N$ and $REB_{28.5}C_4$ (RE = heavy rare earth element, Y, Sc). Recent discovery of B-rich compounds has received considerable research attention for the betterment of the physical properties in these days.[1–3,5,7–14] Very recently, Cherednichenko et al.[8] have synthesized 3D B-rich compounds $B_{12}S$ and $B_{12}Se$ using chemical reaction process at a pressure of 6 GPa and temperature 2500 K. With the help of powder X-ray diffraction and Raman spectroscopy, it is found that the crystal structure belongs to rhombohedral symmetry with a space group of *R-3m* (# No. 166). The stoichiometric ratio of the chalcogenides has been computed. Furthermore, only the bulk modulus of $B_{12}Se$ among all mechanical properties has been studied using third-order Birch-Murnaghan equation of state so far.[7]

The structure of $B_{12}$-icosahedral boron network is also found in some other compounds similar to the $B_{12}S$ and $B_{12}Se$. For instance, the chalcogenide compounds of $B_6S$ and $B_6Se$ have already been synthesized by chemical reaction process at high-pressure. A comprehensive theoretical study of structural, electronic, mechanical, optical and thermal properties on these two compounds have been reported recently which demonstrated their potential for variety of thermo-mechanical applications[1,6]. The boron-rich compounds $B_{12}C_3$[9,10], $B_{12}O_2$[10,11] and $B_{12}As_2$[2] were also reported as very hard compounds with good mechanical strength, superior mechanical stability and the ability to function in harsh environments. Furthermore, Korozlu et al.[12] and Pan et al.[13] have reported the mechanical properties and in particular, the hardness values of some B-rich compounds using various models. They concluded that most of the B-rich chalcogenide compounds are usually prominent members of hard and superhard family. Since the structure of boron-rich $B_{12}$-icosahedral supports the unusual physical and chemical properties, a better knowledge especially related to the physical properties of $B_{12}S$ and $B_{12}Se$ could help to design new applications oriented tasks.

However, many decisive properties of $B_{12}S$ and $B_{12}Se$ compounds like electronic (band structure, density of states, charge density, Mulliken population analysis), optical (dielectric function, absorption coefficient, photoconductivity, reflectivity, loss function and refractive index) and mechanical (elastic stiffness constant, polycrystalline moduli, hardness, fracture toughness, machinability index, melting point etc.) are still unexplored. Understanding these properties is a matter of prime interest in order to disclose the full potential of the compounds of interest for possible device applications.

In the present contribution, structural, electronic, mechanical and optical properties of newly synthesized B-rich chalcogenide compounds, $B_{12}X$ (X = S, Se) have been investigated meticulously with the help of the first-principles method based on density functional theory (DFT) for the first time. The obtained results confirmed that $B_{12}X$ (X= S, Se) are indirect band gap semiconductors that are likely to be hard materials and can also be used as an optical absorber in photovoltaic device as well as optical waveguide. We have also carried out a comparative analysis of physical properties of $B_{12}X$ (X = S, Se) with other similar B-rich compounds, where available.

## 2. Theoretical methodology

In the present report, we focus on the analysis and discussion of structural, electronic, optical and mechanical properties of newly synthesized B-rich chalcogenide compounds, $B_{12}X$ (X= S, Se). To carry out these tasks, first-principles approach based on state-of-the-art density functional theory (DFT) has been employed.[15,16] The high throughput calculations are implemented in the CAmbridge Serial Total Energy Package (CASTEP) module.[17] The choice of pseudopotential is quite important in the point of view of optimization of crystal structure and its electronic structure, especially of the semiconductor materials.[6,18,19] The exchange-correlation potentials are evaluated by using the functional form of Perdew-Burke-Ernzerh of (PBE) type within the generalized gradient approximation (GGA) and also of Ceperly and Alder-Perdew and Zunger (CA-PZ) type within the local density approximation (LDA).[16,20,21] The optimizations for both chalcogenide crystal structures are done by the Broyden-Fletcher-Goldfarb-Shanno (BFGS) method[22] using the following optimization input parameters: plane wave basis set kinetic energy cut-off of 550 eV; Monkhorst–Pack $k$-point mesh size[23] of 6×6×3; energy convergence threshold of 5×10$^{-6}$ eV/atom; maximum force of 0.01 eV/Å; maximum stress of 0.02 GPa; maximum atomic displacement of 5×10$^{-4}$ Å. Very often, the band gap estimation of semiconductor materials using local functional like LDA, GGA are not matched with experimental results and underestimate the actual values. The electronic band gap of both title compounds, therefore, are calculated as accurately as possible by using the non local hybrid functional HSE06 (Heyd–Scuseria–Ernzerhof).[24–26] Elastic stiffness constants can be found by the 'stress-strain' method supported by the CASTEP code.[27] The optical properties such as real part of the dielectric function [$\varepsilon_1(\omega)$], refractive index, absorption spectrum, loss-function, reflectivity and optical conductivity can be estimated following the Kramers-Kronig transformation relation from the imaginary part of the dielectric function, [$\varepsilon_2(\omega)$]. The following equations are used to estimate the aforementioned optical constants where the symbols bear the usual significances[17,27,28]:

$$\varepsilon_2(\omega) = \frac{2e^2\pi}{\Omega\varepsilon_0} \sum_{k,v,c} |\psi_k^c|\boldsymbol{u}.\boldsymbol{r}|\psi_k^v|^2 \delta(E_k^c - E_k^v - E)$$

$$\varepsilon_1(\omega) = 1 + \frac{2}{\pi} P \int_0^\infty \frac{\omega' \varepsilon_2(\omega') d\omega'}{(\omega'^2 - \omega^2)}$$

$$n(\omega) = \frac{1}{\sqrt{2}} \left[ \sqrt{\{\varepsilon_1(\omega)\}^2 + \{\varepsilon_2(\omega)\}^2} + \varepsilon_1(\omega) \right]^{\frac{1}{2}}$$

$$\alpha(\omega) = \sqrt{2}\,\omega \left[ \sqrt{\{\varepsilon_1(\omega)\}^2 + \{\varepsilon_2(\omega)\}^2} - \varepsilon_1(\omega) \right]^{\frac{1}{2}}$$

$$L(\omega) = \varepsilon_2(\omega)/[\{\varepsilon_1(\omega)\}^2 + \{\varepsilon_2(\omega)\}^2]$$

$$R(\omega) = \left| \frac{\sqrt{\varepsilon(\omega)} - 1}{\sqrt{\varepsilon(\omega)} + 1} \right|^2$$

3. **Results and discussion**

   3.1. **Structural properties**

The crystal structure of two B-rich chalcogenide compounds $B_{12}S$ and $B_{12}Se$ are isostructural to α-rhombohedral boron (α-$B_{12}$). The titled compounds contain 10 (ten) $B_{12}$-icosahedra (a three-dimensional polyhedral rigid network) in the hexagonal unit cell in which eight icosahedra unit are found at corner positions and the other two are placed on one of the main diagonals.[8] The crystallographic lattice parameters of both chalcogenides under study are presented in Table-1, along with previously reported data. The optimized crystal structure of $B_{12}S$ as prototype structure is depicted in Fig.1. It is obvious that the lattice constants and bond lengths of both studied compounds could be different due to the variation of electronegativity and atomic number of 6c site element (S/Se). In Table-1, there have been some minor variations of both lattice constants ($a, c$) in the literature data due to the variation of S/Se occupancy in the unit cell. It was reported that the level of occupancy of S atom at 6c site is 55% and that of Se at 6c site is to be 52%, while all the boron atoms form $B_{12}$-icosahedra having occupancies at 18h sites are found to be 100%. These results reveal an excellent final reliability factor of the Rietveld refinement analysis as well as of the energy-dispersive X-ray spectroscopy data.[8] However, the calculated lattice parameters using LDA-CAPZ functional give fair agreement with other literature data compared to those obtained via GGA-PBE.

**Table 1:** Optimized crystallographic lattice parameters, $a$ and $c$ (all in Å), unit cell volume $V$ ($Å^3$) of boron sub-sulfide ($B_{12}$S) and boron sub-selenide ($B_{12}$Se) compounds along with prior literature data.

| Phase | $a$ | $c$ | $V$ | Occupancy (%) | Functional | Ref. |
|---|---|---|---|---|---|---|
| $B_{12}$S | 5.8392 | 12.1115 | 357.6302 | 55 | LDA-CAPZ | This |
|  | 5.9118 | 12.3155 | 372.7574 | 55 | GGA-PBE | This |
|  | 5.8196 | 11.9653 | 364.76 | 55 | Exp. | 8 |
|  | 5.8966 | 12.1135 |  | 100 |  | 8 |
|  | 5.80 | 11.90 |  | 50 |  | 29 |
|  | 5.810 | 11.94 |  | 48.5 |  | 30 |
|  | 5.8624 | 12.147 |  | 65 |  | 30 |
|  | 5.8379 | 12.036 |  | 59.9 |  | 14 |
|  | 5.8307 | 12.028 |  | 60.9 |  | 14 |
|  | 5.8273 | 12.025 |  | 62 |  | 14 |
| $B_{12}$Se | 6.0237 | 12.1596 | 382.1013 | 52 | LDA-CAPZ | This |
|  | 6.0786 | 12.3713 | 395.8673 | 52 | GGA-PBE | This |
|  | 5.9385 | 11.9144 | 385.42 | 52 | Exp. | 8 |
|  | 6.0496 | 12.1603 |  | 100 |  | 8 |
|  | 5.9041 | 11.947 |  | 46.9 |  | 5 |

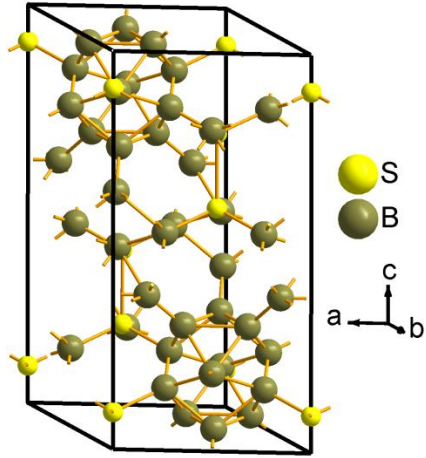

**Fig. 1.** Schematic crystal structure of $B_{12}S$ chalcogenide compound belonging to the rhombohedral symmetry with space group of *R-3m* (# No. 166). Two icosahedra units are found on one of the main diagonals. The $B_{12}Se$ chalcogenide compound is completely isostructural to $B_{12}S$ and is not shown here.

### 3.2. Electronic properties, charge density and Mulliken population

The electronic band structure (EBS) along high symmetry directions of $B_{12}S$ and $B_{12}Se$ are illustrated in Fig. 2.The Fermi level ($E_F$) is set to 0 eV. In many semiconducting materials, band gap estimation using local functional results in inaccurate value along with inappropriate positions of the valence band (VB) and conduction band (CB), which lead the theoretical study far-off from the experimental results. The error might emanate from strong Coulomb correlations of the material. In most cases, the use of non-local functional could be a fair solution of the aforementioned problem.[25,26] In order to provide an accurate prediction of EBS as well as the band gap we have used both local functional (LDA and GGA) and non-local functional (HSE06) to model the exchange-correlation potential. At present, due to the lack of prior literature data, comparison of EBS results with any experimental and/or theoretical one is not possible. All the results presented in this section will require experimental validation. However, as depicted from Fig. 2, the bottom of the CB is at fixed position at A-point of the Brillouin zone for both the compounds while the top of the VB are located at different places; M-point and Γ-point for $B_{12}S$ and $B_{12}Se$, respectively. This implies that the both the chalcogenides should be indirect band gap semiconductors.

The value of indirect band gap is very sensitive to the exchange and correlation functional as shown in Fig. 2. The values of indirect band gaps (in eV) for $B_{12}S$ ($B_{12}Se$) compounds using LDA, GGA and HSE06 functionals are found to be 1.287 (0.0), 1.245 (0.093) and 2.271 (1.300), while that of minimum direct band gap at M-point (Γ-point) are 1.985 (1.495), 2.161 (1.564) and 2.746 (2.344), respectively. The band morphology was significantly affected by the chalcogenide element (X element). For instance, due to the replacement of S element with Se in $B_{12}S$ compound, all valence and conduction bands (in case of use of both local and non-local functionals) are shifted towards the $E_F$ with minor variation in the shape of the bands. Similar results were also reported in the B-rich chalcogenide, $B_6X$ (X= S, Se) compounds.[6] Furthermore, the top of the VB was shifted from M-point to Γ-point when the S element is substituted by Se atom to form the compound of $B_{12}Se$. It is also seen that the band along H-K and H-L directions for both compounds are much less dispersive than in any other direction. Furthermore, the bottom of the CB and the top of the VB based on the band morphology inspection inspire us to calculate the effective mass which can predict the charge transport behaviors. The effective mass of electron $(m_e^*)$ and hole $(m_h^*)$ for CB and VB have been calculated, respectively, by the following equation[31]:

$$m^* = \frac{\hbar^2}{\left(\frac{d^2\mathcal{E}(k)}{dk^2}\right)}$$ where, $k$ is the wave vector, $\mathcal{E}(k)$ is the eigenvalue of the energy band at wave vector, $k$ and $\hbar$ is the reduced Planck constant. It is obvious from the equation that the value of $m^*$ has an inverse relationship with the curvature of the electronic band dispersions. The calculated carrier effective mass ($m_e^*$m, $m_h^*$m), average effective mass ($m_{dc}$, $m_{dv}$) and band gap (eV) for boron sub-sulfide ($B_{12}S$) and boron sub-selenide ($B_{12}Se$) compounds are displayed in Table 2. The average effective mass of $m_{dc}$ and $m_{dv}$ at CB and VB, respectively are calculated using standard formulae that can be found elsewhere.[31,32]

The calculated effective masses of electron $(m_e^*)$ at A-point along (A→Γ) and (A→H) directions in the CB for both compounds are different due to the difference in curvatures of the bands. Similar results are also found for hole in the VB as shown in Table 2. In $B_{12}Se$ compound, the $m_{dc}$ has a light effective mass compared to that of $B_{12}S$ compound, which originates from dispersive $p$ orbitals of the B element in the former. In contrary to $m_{dc}$, the $m_{dv}$ with value of $3.37m_0$ at Γ-point for $B_{12}Se$ compound are much heavier than that $1.94m_0$ at M-point for the $B_{12}S$ compound. Here, $m_0$ is the bare electron mass. This indicates that less-dispersive curvature (almost flat band) is formed at Γ-point in comparison to that at M-point. Noted here that the top of VB was shifted from M- to Γ-point with the replacement of the S atom with Se element. It is known that the effective mass can predict the carrier mobility as well as carrier concentration of a semiconductor material and are often used to predict electrical conductivity of solid. A low value of $m_{dc}$ for $B_{12}Se$ should indicate high carrier mobility as well as electrical conductivity under external stimuli.

In order to extend the nature of the EBS, we have also studied the electronic energy density of states (total and partial density of states) of the titled compounds as shown in Fig. 3. Considering the hybrid functional of HSE06, the lowest energy parts of the CB for $B_{12}S$ ($B_{12}Se$) at 2.30 (1.30) eV are formed by the strong hybridization of $s$ and $p$ orbitals of B element, while hybridization between $s$ and $p$ orbitals of S element has minor contribution. On the other hand, the highest energy parts of the VB in the vicinity of $E_F$ is mainly constructed by the hybridization of $p$ orbitals of both B and S/Se elements.

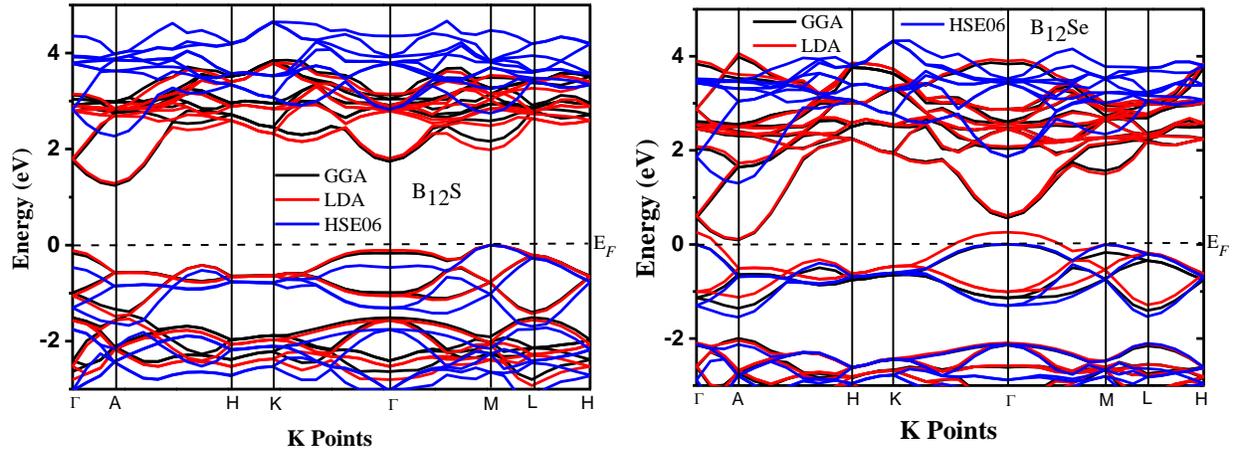

**Fig. 2.** Electronic band structures of boron-rich chalcogenide compounds $B_{12}S$ and $B_{12}Se$ along the high symmetry directions using local (GGA and LDA) and non-local (HSE06) functionals. The black dashed horizontal lines denote the Fermi level ($E_F$), which are set to 0 eV.

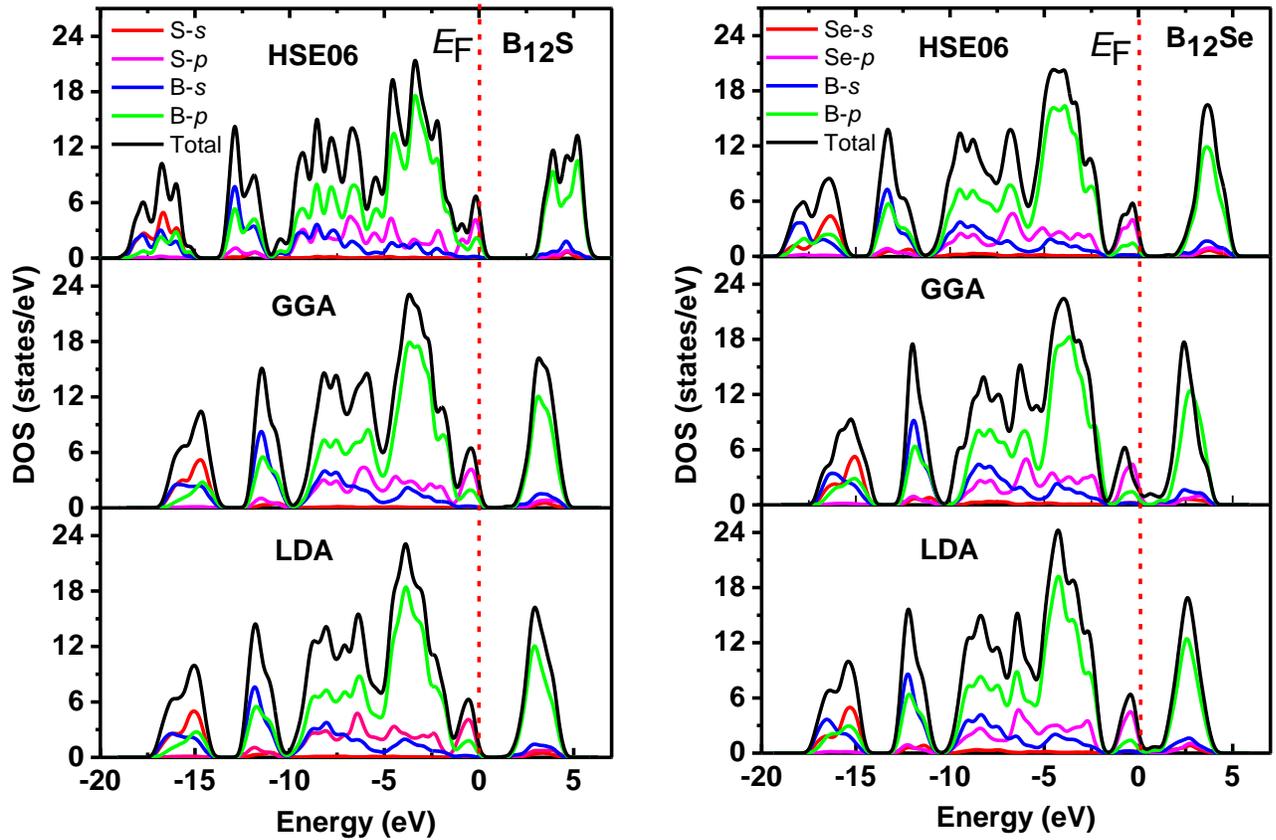

**Fig. 3.** The density of states (total and partial) of boron-rich chalcogenide compounds $B_{12}S$ and $B_{12}Se$ using LDA, GGA and HSE06 functionals. The red dashed vertical lines denote the Fermi level ($E_F$) set to 0 eV.

It can be seen here from the PDOS plots that hybridization of B and Se-$p$ orbitals are much stronger than that between B and S-orbitals, resulting in shift of lowest conduction band downwards (lower energy). Similar qualitative features could be found for both the titled chalcogenides using LDA and GGA functionals to model the exchange and correlation effects.

**Table 2:** Carrier effective mass ($m_e^*$, $m_h^*$), average effective mass ($m_{dc}$, $m_{dv}$) and band gap (eV) for boron sub-sulfide ($B_{12}S$) and boron sub-selenide ($B_{12}Se$) compounds.

| Parameters/ Approach | $B_{12}S$ | $B_{12}Se$ | Ref. |
|---|---|---|---|
| $m_e^*(m_0)$ | 0.98 A(A→Γ) <br> 1.07 A(A→H) | 0.96 A(A→Γ) <br> 0.59 A(A→H) | This |
| $m_{dc}(m_0)$ | 1.02 | 0.75 | This |
| $m_h^*(m_0)$ | 1.85 M(M→Γ) <br> 2.04 M(M→L) | 3.50 (Γ→M) <br> 3.24 Γ (Γ→M) | This |
| $m_{dv}(m_0)$ | 1.94 | 3.37 | This |
| Bandgap(eV)(LDA-CAPZ) | 1.287 | 0 | This |
| Bandgap(eV) (GGA-PBE) | 1.245 | 0.093 | This |
| Bandgap(eV) (HSE06) | 2.271 | 1.300 | This |

**Table 3:** Estimated Mulliken atomic population of boron-rich chalcogenides, $B_{12}X$ (X= S, Se). Here, EVC indicates the effective valence charge.

| Compound | Atom | s | p | Total | Charge (e) | EVC (e) |
|---|---|---|---|---|---|---|
| $B_{12}S$ | B | 0.83 | 2.17 | 3.00 | 0.00 | --- |
|  | B | 0.79 | 2.29 | 3.08 | -0.08 | --- |
|  | S | 1.59 | 4.17 | 5.76 | 0.24 | 5.76 |
| $B_{12}Se$ | B | 0.88 | 2.22 | 3.10 | -0.10 | --- |
|  | B | 0.85 | 2.36 | 3.22 | -0.22 | --- |
|  | Se | 0.97 | 4.08 | 5.05 | 0.95 | 5.05 |

The three dimensional (3D) mapping image of charge density distribution of $B_{12}S$ and $B_{12}Se$ are depicted in Fig. 4. It can be deduced from the figures that there are asymmetric charge distributions in the studied compounds owing to the difference in the electronegativity of S and Se atomic species. The depletion of charge density indicated by blue color is found around S/Se atoms while the accumulation of charge is pointed out by red color as seen around the B atomic species. This reflects that charge (electron) is shifted from S/Se element to B element, which is suggestive of ionic bonding. In addition to this, strong covalent bonds between B – B and B – S/Se atoms are found owing to the directional accumulation of charge. The charge density distributions in both studied compounds are different as seen from charge accumulation and depletion regions. In the $B_{12}Se$ compound, charge accumulation around B atoms and depletion around Se element is more pronounced than that in the $B_{12}S$ compound. This suggests that amount of charge transfer as well as bonding strength could be different in these boron-rich compounds[33] which is also confirmed from the study of Mulliken population analysis as shown in Table 3. The B and S/Se atoms have a negative and positive charge, respectively. The transfer of charge from S to B atoms in $B_{12}S$ compound is $0.08|e|$ while in $B_{12}Se$ compound, it is from Se to B atoms and is $0.22|e|$ and $0.10|e|$ as indicated in Table 3. These results are consistent with the results of charge density distribution mapping.

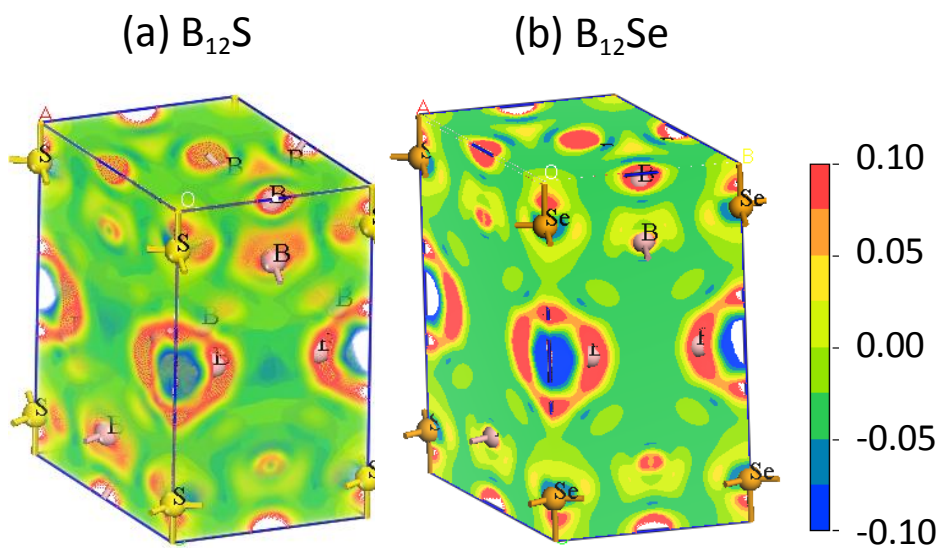

**Fig. 4.** The 3D mapping image of charge density distribution of the titled chalcogenides: (a) $B_{12}S$ and (b) $B_{12}Se$ compounds with adjacent scale showing the charge value (in the units of electronic charge) from -0.1 to +0.1.

### 3.3. Optical properties

In order to judge the feasibility of a material for possible optoelectronic and photonic device applications as well as for the characterization of materials, especially semiconducting properties, it is of paramount importance to study various optical features, at first. Generally, optical property is related to a material's response when electromagnetic radiation, particularly visible light is exposed on it. Some important optical devices such as LCD screens, screens for 3D movies, polarizers, wave plates, to name a few, are fabricated based on the properties of optical anisotropy.[6,18,34] Therefore, the optical properties along with anisotropy in a broad spectral range are of decisive interest to extend their application scope. In present investigation, the optical properties as well as optical anisotropy of two titled compounds are studied in the photon energy range up to 15 eV for two polarization directions [100] and [001], for the first time. It is interesting to note here that the shapes of the spectrum and energy peak position for all optical properties are different and distinguishable with respect to the polarization direction of the incident electric field. The direction-dependent real [$\varepsilon_1(\omega)$] and imaginary [$\varepsilon_2(\omega)$] parts of the dielectric function for both compounds are illustrated in Fig. 5. It is seen that large anisotropic behavior of both dielectric functions is observed in the IR, visible and mid-UV region up to 12 eV. Particularly, very strong anisotropic nature of these dielectric functions in the IR and visible region are crucially important for efficient light manipulation in many devices. Following the Penn model[35,36], the static value of dielectric constant [$\varepsilon_1(0)$] can be evaluated from the plasma energy ($E_p$), Fermi energy ($E_F$), and energy bandgap ($E_g$) as follows:

$$\varepsilon_1(0) = 1 + \left(\frac{E_p}{E_g}\right)^2 \left[1 - \frac{E_g}{4E_F} + \frac{1}{3}\left(\frac{E_g}{4E_F}\right)^2\right]$$

The static value of [$\varepsilon_1(0)$] that describes the index of refraction, is also necessary to design many optical devices. For $B_{12}Se$, the highest values of [$\varepsilon_1(0)$] is calculated using GGA (HSE06) functional and found to be 5.08 (3.51) and 4.75 (3.23) for [100] and [001] polarizations, respectively, while the corresponding values are 4.75 (3.41) and 4.07 (3.05) of $B_{12}S$ for [100] and [001] polarization directions, respectively. The [$\varepsilon_1(\omega)$] for both the compounds using GGA (HSE06) functional are gradually increased with photon energy, reaching the highest values of 7.75 (5.36) at 5.2 (6.2) eV for $B_{12}S$ and 8.20 (6.04) at 5.22 (6.16) eVf or $B_{12}Se$ for [100] direction and finally goes to zero at around 6.5 (7.5) eV energy. The photon energy/frequency dependent imaginary part of the dielectric constant [$\varepsilon_2(\omega)$] signifies the light absorption of the material. The values of [$\varepsilon_2(\omega)$] for $B_{12}S$ using GGA (HSE06) functional started at incident photon energy of 1.9 (2.56) eV and 1.8 (2.77) eV along [100] and [001] directions, respectively, and for $B_{12}Se$ the values are 1.35 (2.40) eV and 1.0 (2.30) eV for the same. The major peaks observed in [$\varepsilon_2(\omega)$] for both chalcogenides arise due to the optical transition mainly from the highest energy of the valence band of S/Se-$p$ orbitals to the lowest energy of the conduction band of B-$p$ orbitals. The positions of prime peaks are varied along different polarization directions with photon energy which is a clear signature of the optical anisotropy, as depicted in Fig. 5.

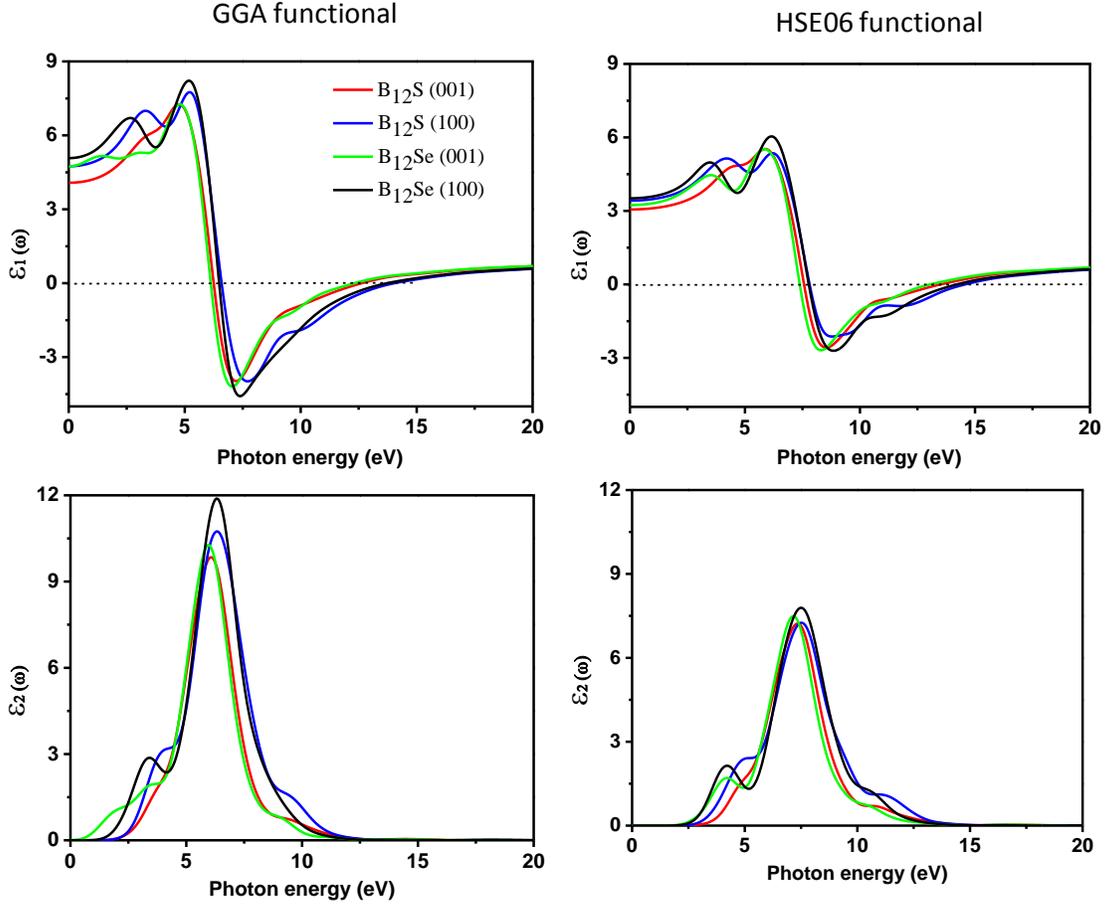

**Fig. 5.** Photon energy dependent dielectric functions (real part [$\varepsilon_1(\omega)$] and imaginary part [$\varepsilon_2(\omega)$]) of boron sub-sulfide ($B_{12}S$) and boron sub-selenide ($B_{12}Se$) compounds along different polarization directions of electric fields using the GGA and hybrid HSE06 functionals.

The optical absorption coefficient, $\alpha(\omega)$ can be computed using the formula mentioned in the theoretical methodology (Section 2). Fig. 6 shows the energy dependent absorption coefficient curves for two polarization directions of [100] and [001]. It is noticed that the $\alpha(\omega)$ revealed strong anisotropy along [100] and [001] polarization directions and in particular, large anisotropy in absorption coefficient is observed in visible spectrum and UV (up to ~ 5.0 eV) and then anisotropy is drastically reduced at an energy around 8 eV. Interestingly, bulk absorption anisotropy is again observed in the deep UV region within an energy range of 8.0-14.5 eV. Since visible light absorption coefficient as well as anisotropy is very important to design an optical device[34], we are more interested to put deep insight in this aspect. Fig. 7 shows the enlarged spectrum in the visible region. It is obvious that the spectrum of $\alpha(\omega)$ along [100] directions has a blue shift in comparison to that along [001] polarization directions for both the boron-rich semiconductors. This implies that strong visible-light absorption should be found along in-plane direction.

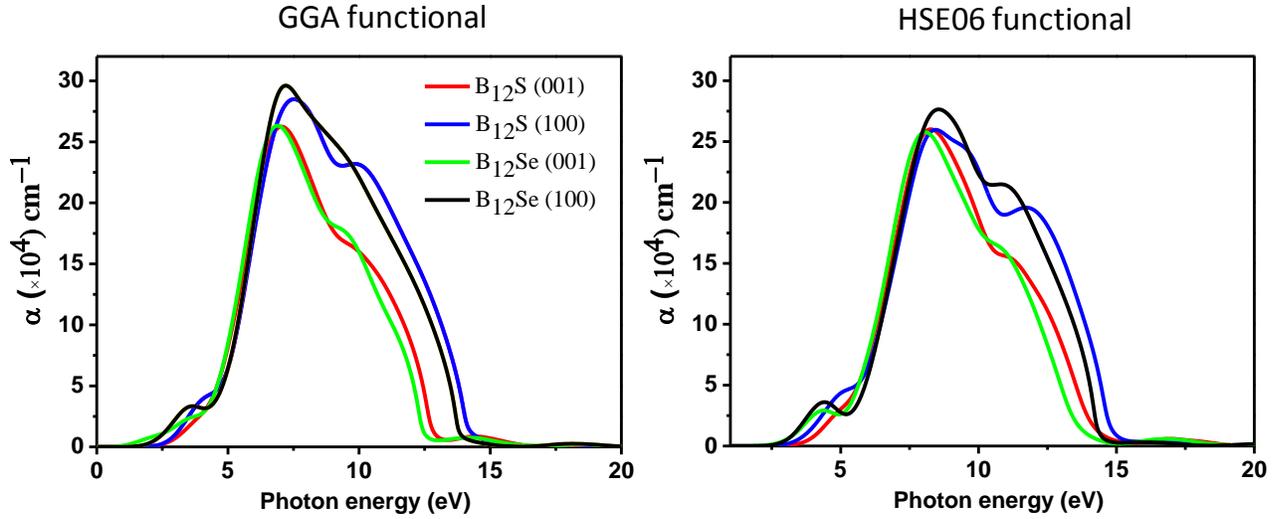

**Fig. 6.** The computed optical absorption coefficient, α(ω) of boron sub-sulfide ($B_{12}S$) and boron sub-selenide ($B_{12}Se$) compounds for different polarization directions of the electric field using GGA and hybrid HSE06 functionals.

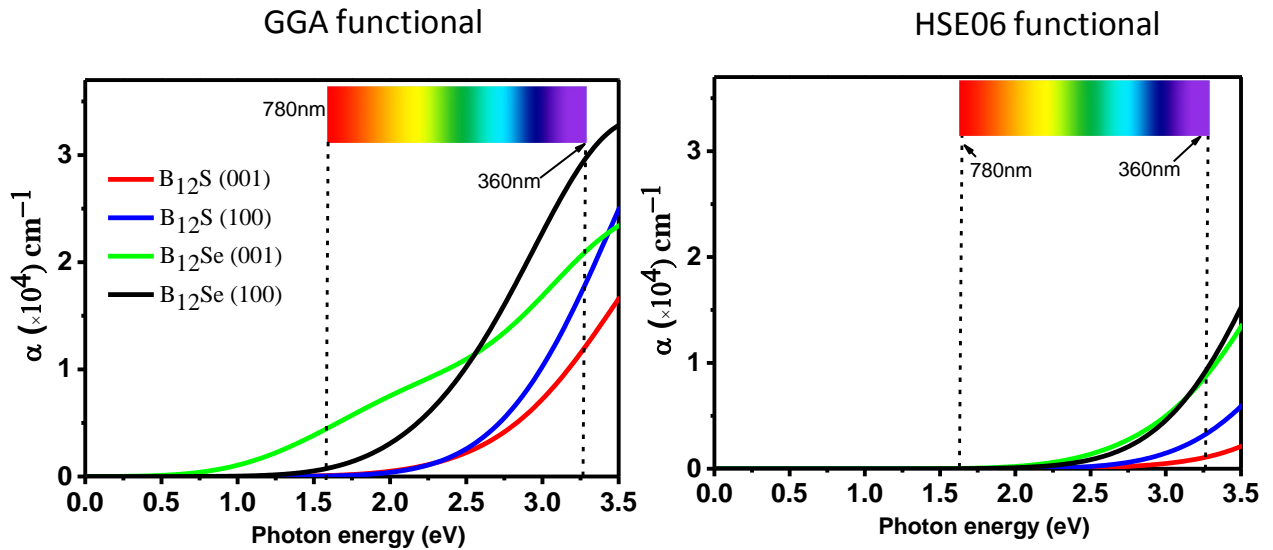

**Fig. 7.** Enlarged figure of absorption coefficient, α(ω) spectrum in the visible spectrum for $B_{12}S$ and $B_{12}Se$ compounds using GGA and hybrid HSE06 functionals.

Therefore, the boron sub-sulfide ($B_{12}S$) and boron sub-selenide ($B_{12}Se$) compounds as an optical absorber in photovoltaic device, for example, should be used parallel to optical source in order to achieve the best absorption efficiency.[33] The major peaks of α(ω) using GGA (HSE06) functional are noticed at 7.56 (8.44) eV and 7.00 (8.26) eV for $B_{12}S$, and that are at (7.16) 8.56 eV and (7.00) 8.06 eV for $B_{12}Se$ compound for [100] and [001] directions, respectively. The maximum value of α(ω) in visible light region using GGA (HSE06) functional is ~$3\times10^4$ ($1\times10^4$) $cm^{-1}$ for $B_{12}Se$ compound.

The optical band gap can also be determined for both the studied chalcogenides using the well-known Tauc plot from absorption coefficient spectra.[37] An accurate estimation of optical band gap energy is quite important to predict some features such as photo-physical and photo-chemical properties of semiconducting materials. On the other hand, a wrong use of Tauc plot to estimate this band gap energy provides misinformation and in particular, error in estimation of band gap can reflect the light absorption within the sub-band gap energy.[38] However, the Tauc equation which is based on the absorption coefficient can be written as follows: $(\alpha.h\nu)^{\frac{1}{\eta}} = A(h\nu - E_g)$ where, the $\alpha, h, \nu, A$ and $E_g$ signify the absorption coefficient, Planck constant, frequency, a material dependent coefficient and the band gap, respectively.[37] The value of parameter $\eta$ depends on nature of band gap and in general, ½ and 2 are used for direct and indirect band gap semiconductors, respectively. Fig. 8 illustrates the optical band gap energy estimation using the Tauc plot for both the chalcogenides under consideration. It is found that the obtained optical band gaps for $B_{12}S$ and $B_{12}Se$ compounds are 2.19 (2.33) eV, and 1.52 (1.32) eV, respectively for [100] ([001]) calculated using the HSE06 functional. The optical band gaps are consistent with those obtained from the electronic band structure gaps as shown in Fig. 2 and Table2.

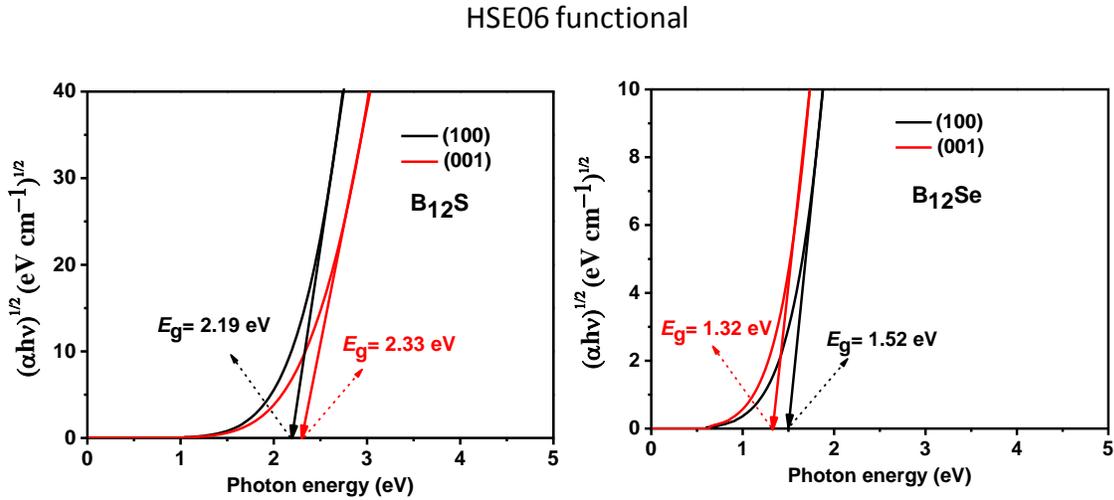

**Fig. 8.** Estimation of optical band gap from the absorption coefficient, α(ω), spectrum using the Tauc plot for $B_{12}S$ and $B_{12}Se$ compounds using the hybrid HSE06 functional.

The refractive indices are also studied to further check the suitability of the two compounds in optoelectronic devices. When a light wave (photons) penetrates through a material, its electric field interacts with the electrons of solids and the real part of the refractive index measures how the velocity of light wave is modified compared to its free-space value. It is seen in Fig. 9 that the value of *n* is also dependent on the polarization directions of the incident electric fields as the static value of *n*(0) at zero frequency using GGA(HSE06) functional along [100]([001]) direction is 2.17(1.84) and 2.01(1.75) for $B_{12}S$, and 2.25(1.87) and 2.17(1.80) for $B_{12}Se$, respectively. Since the value of *n* is different in different directions, the effective refractive index must be dependent on electric field direction of the light. However, the variations of these values in the visible light region are not that prominent; rather reasonably constant. Electric polarization dependent value of *n* also suggests that higher value of *n*[100] results due to the highest density of ions along these directions.[34] The value of *n* considering HSE06 functional is lower than that of $B_6X$ (X = S and Se)[6], $NaInSe_2$[18], MAX[39,40], MAB[41–43] but is slightly higher than those of some ceramic materials such as silica glass, quartz and soda lime glass; and in some compounds, very close to corundum, periclase and spinel[44]. A comparison of *n* (at 0 eV) with some well-known optical waveguide materials is presented in the lower panel of Fig. 9. This comparison implies that the titled compounds are likely to be used as optical guides.

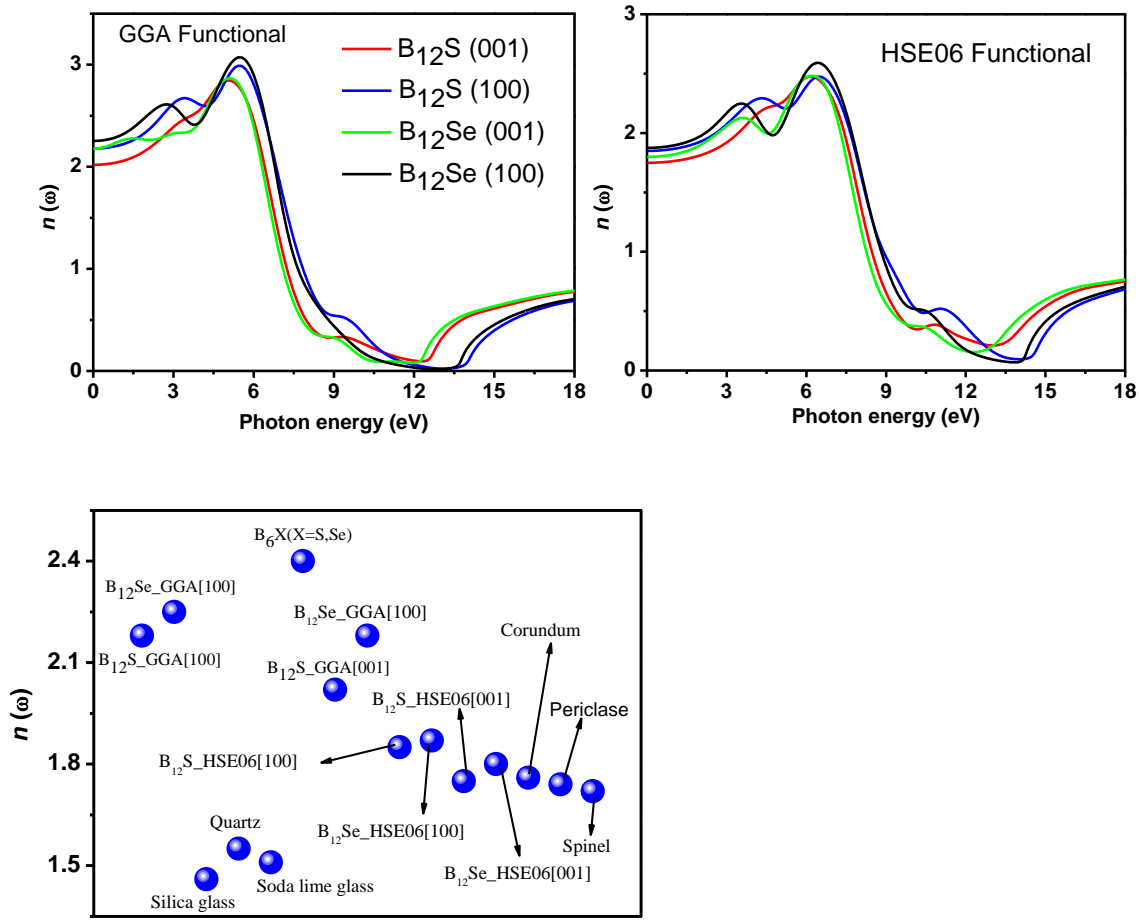

**Fig. 9.** The refractive index spectra, *n* of $B_{12}S$ and $B_{12}Se$ compounds obtained using the GGA and hybrid HSE06 functional as a function of incident photon energy. Lower panel: comparison of static refractive index, *n*(0) of $B_{12}S$ and $B_{12}Se$ with some other well-known ceramic and optical waveguide materials.

The reflectivity spectra of boron sub-sulfide and sub-selenide are estimated and plotted as a function of photon energy in Fig. 10. It is found that the reflectivity spectrum is very sensitive in the IR, visible and UV-light (up to 5.0 eV) regions when the electric field polarization is along [100] and [001] directions. Like the anisotropic behavior of absorption coefficient, the anisotropy of the reflectivity spectra was also drastically reduced in the photon energy range of 5.0 - 8.0 eV and then again at higher energies the anisotropic nature is prominently noticed. The starting values of the spectra for $B_{12}S$ are 0.0888 (0.14) and 0.0742 (0.113) for [100] and [001], respectively calculated using the HSE06 (GGA) functional. On the other hand, the starting values for $B_{12}Se$ are 0.0925 (0.1483) and 0.0815 (0.1375) for the same. Polarization direction dependent maximum reflectivity spectra are seen for both the functional in the following sequence: $B_{12}Se$ [100] > $B_{12}S$ [100] > $B_{12}Se$ [001] > $B_{12}S$ [001] at various incident photon energies as shown in Fig.10.

The transparency $T(\omega)$ of any material can be estimated with the help of absorption coefficient, $\alpha(\omega)$ and reflectivity, $R(\omega)$ data as follows[31]:

$$T(\omega) = (1 - R(\omega))e^{-\alpha(\omega)t}$$

$\omega$ and t are the light frequency and thickness of the material, respectively. Lower values of the absorption coefficient or the reflectivity results in the higher value of the transparency as indicated from the above equation. It is worth noticing from the Figs.10 and 6 that the reflectivity spectra and absorption coefficients of both the compounds have the values of reflectivities and absorption coefficients ~10.0% and ~$1.0\times10^4$ cm$^{-1}$, at least in the visible light region. This indicates that the titled chalcogenide compounds are fairly transparent materials.

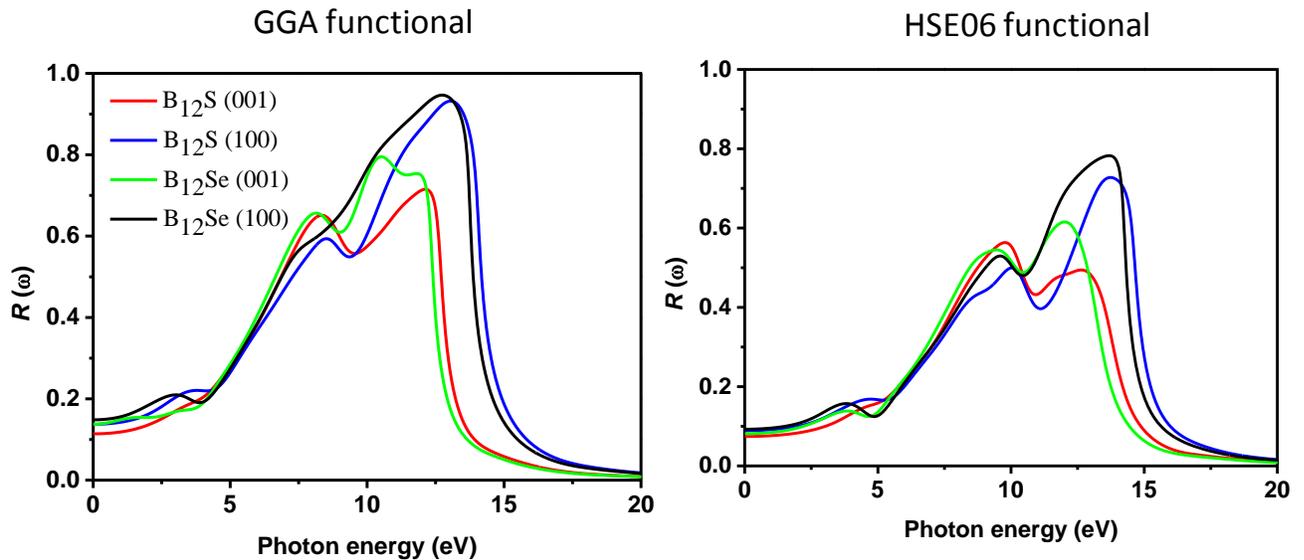

**Fig. 10.** The reflectivity spectra of boron sub-sulfide ($B_{12}S$) and boron sub-selenide ($B_{12}Se$) compounds plotted as a function of photon energy.

The loss function, $L(\omega)$, is related to the loss of energy of charged particles (e.g., electrons) when they travel through a material. As can be shown from Fig. 11 that energy loss of electron is prominently seen in the mid-UV region (within energy range 12- 15 eV for both the compounds). Very sharp peaks are attained at the plasma frequency. The plasma resonances using the GGA (HSE06) functional are at 14.04(14.6) eV and 12.64(13.60) eV for $B_{12}S$ and at 13.72 (14.22) eV and 12.31 (13.11) eV for $B_{12}Se$, along [100] and [001] polarization directions, respectively. Additionally, it can be seen that there have been no $L(\omega)$ peaks in the range 0-12 eV owing to the high value of $[\varepsilon_2(\omega)]$.[45] The imaginary part of the dielectric function is greatly suppressed at the plasma resonance.

It is found that like other optical properties discussed herein, the energy loss function for both the compounds could be higher along with larger plasmon energy when for [100] polarization in comparison to that for [001] direction. Prior to the present study, there has been no experimental and/or theoretical report concerning the optical properties of $B_{12}S$ and $B_{12}Se$ to validate our theoretical estimations. The results reported in this section will require future experimental verification.

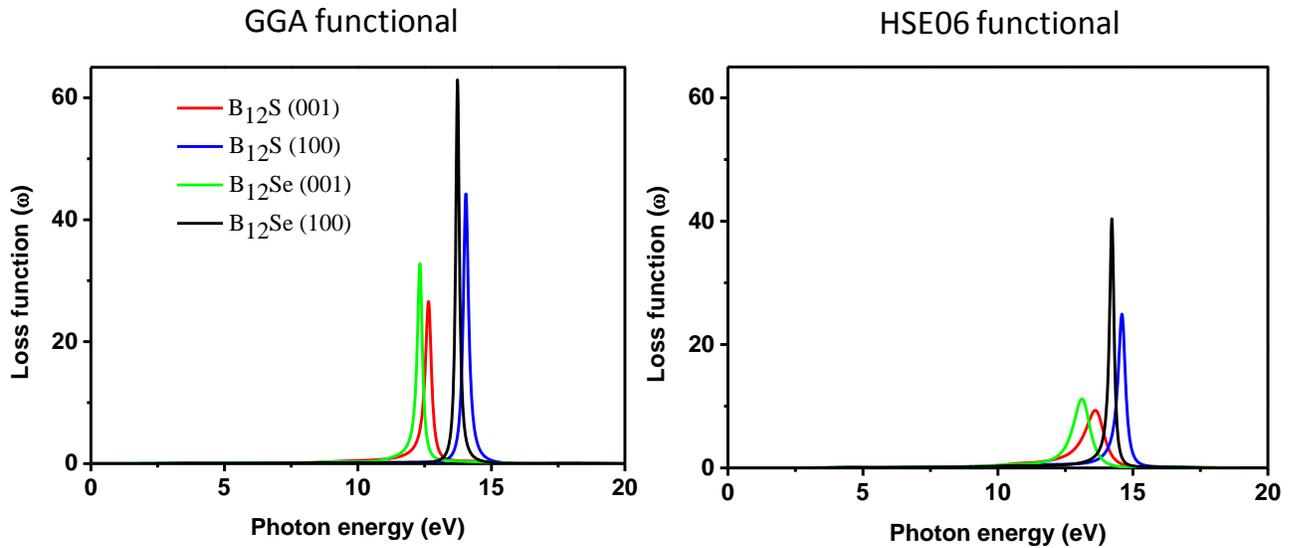

**Fig. 11.** The photon energy dependent loss function of boron sub-sulfide ($B_{12}S$) and boron sub-selenide ($B_{12}Se$) compounds.

### 3.4. Mechanical properties and hardness

In order to comprehend the mechanical behaviors of the two compounds, we have calculated the mechanical properties characterizing parameters such as elastic constants ($C_{ij}$), elastic moduli, hardness parameters, brittle/ductile behavior, etc. The calculation of elastic constants $C_{ij}$ via CASTEP code was performed using efficient stress-strain method.[46,47] The stress-strain relationship is usually expressed by the generalized Hooke's law: $\sigma_{ij} = C_{ijkl}\varepsilon_{kl}$. Here, the fourth-order elastic stiffness tensor $C_{ijkl}$ contains 81 components, $\sigma_{ij}$ and $\varepsilon_{kl}$ are the second-order stress tensor and strain tensor, respectively. The 6 × 6 matrix can be used to express the fourth order tensor using the Voigt markers and the equation can be simplified to: $\sigma_i = C_{ij}\varepsilon_j$, as $C_{ij} = C_{ji}$, the number of independent components in the elastic stiffness tensor can be reduced to 21. Due to the symmetry of the crystal structure, the number of single crystal elastic constants is different for different crystal systems.

**Table 4:** Calculated stiffness constants (GPa), elastic moduli (*B, G & Y*; all in GPa), Poisson's ratio (υ), Pugh's ratio (*G/B*), hardness values ($H_{Chen}$, $H_{Miao}$ and $H_{Mazhnik}$; all in GPa) and fracture toughness, $K_{IC}$, (MPam$^{0.5}$) of boron-rich chalcogenide, $B_{12}X$ (X = S, Se) compounds with hexagonal structure.

| Phase | | $C_{11}$ | $C_{33}$ | $C_{44}$ | $C_{12}$ | $C_{13}$ | $C_{14}$ | B | G | Y | υ | G/B | $H_{Chen}$ | $H_{Miao}$ | $H_{Mazhnik}$ | $K_{IC}$ |
|---|---|---|---|---|---|---|---|---|---|---|---|---|---|---|---|---|
| $B_{12}S$ | GGA | 463 | 352 | 119 | 108 | 54 | 30 | 187 | 148 | 351 | 0.18 | 0.79 | 25.29 | 30.89 | 20.32 | 4.27 |
| | LDA | 487 | 371 | 121 | 124 | 61 | 35 | 201 | 151 | 362 | 0.19 | 0.75 | 23.93 | 30.24 | 19.27 | 4.55 |
| $B_{12}Se$ | GGA | 386 | 313 | 86 | 151 | 46 | 56 | 171 | 096 | 243 | 0.26 | 0.56 | 11.69 | 15.13 | 11.57 | 3.13 |
| | LDA | 437 | 336 | 99 | 134 | 51 | 36 | 183 | 126 | 307 | 0.22 | 0.68 | 18.88 | 23.52 | 14.56 | 3.80 |

Since neither experimental nor theoretical data of elastic properties of both the titled phases is available, thus, a comparison between the values obtained for different functionals can be useful to predict the reliability of our calculations. The values of elastic constants and elastic moduli show opposite trend to that of the lattice constants i.e., LDA (GGA) gives lower (higher) lattice constants that result in higher (lower) elastic parameters, as expected. It is also proved that LDA and GGA give higher and lower values of elastic constants, respectively[48]. This inspires us to calculate the elastic properties using these functionals with the hope that the exact values of the elastic parameters for $B_{12}S$ and $B_{12}Se$ should remain within this limit. Table 4 enlists the obtained parameters used to describe the mechanical behaviors of the considered compounds. One of the decisive information that can be revealed from the obtained elastic constants ($C_{ij}$) is the confirmation of the mechanical stability of the studied materials. The conditions of mechanical stability for hexagonal systems are : $C_{11} > 0$, $C_{11} > C_{12}$, $C_{44} > 0$, $(C_{11} + C_{12})C_{33} - 2(C_{13})^2 > 0$.[49] It is obvious from Table 4 that the obtained values of $C_{ij}$ satisfy the mentioned criteria and hence both the compounds are mechanically stable. Some more useful information can be extracted from the values of $C_{ij}$. As evident from Table 4 that $C_{11}$ is larger than that of $C_{33}$ for both functionals. As $C_{11}$ and $C_{33}$ are directly related to atomic bonds along $x(y)$ and $z$ axes, respectively, the atoms along $x(y)$ axis are bonded strongly compared to that along the $z$ axis for both the phases. Table 4 reveals that $C_{11}$ and $C_{33}$ of $B_{12}S$ are larger than those of $B_{12}Se$, indicating higher resistance to axial deformation of $B_{12}S$ in comparison with $B_{12}Se$. Another elastic constant, $C_{44}$, usually used to measure the shear deformation resistance, is also much higher for $B_{12}S$ than that of $B_{12}Se$, implying higher shear deformation resistance of $B_{12}S$ compared to $B_{12}Se$. The unequal values of $C_{11}$ and $C_{33}$ also reveal the difference in the atomic arrangements along $x(y)$ and $z$ axes. Such arrangement of atoms is desired for hexagonal system to ensure the minimum energy of the system in the ground state; consequently, density of atoms along $x(y)$ and $z$ axes is different that results in different values of $C_{11}$ and $C_{33}$. In addition, $C_{11}$ and $C_{33}$ are larger than those of $C_{12}$, $C_{13}$ and $C_{14}$, implying that shear deformation is easier than that of axial deformation for both the boron-rich phases. Moreover, the unequal values of $C_{11}$ and $C_{33}$, and $C_{12}$, $C_{13}$ and $C_{14}$ reveal the anisotropy in the elastic moduli ($Y$, $B$, and $G$). Finally, the elastic constants are used to calculate the macroscopic elastic moduli and ratios such as bulk modulus ($B$), shear modulus ($G$), Young's modulus ($Y$) and Poisson's ratio ($v$) using the Voigt-Reuss-Hill (VRH) approximation.[50–52] The required expressions for these calculations can be found elsewhere.[6,53,54] The obtained values are listed in Table 4 from which it is seen that the $B$, $G$ and $Y$ are 187 GPa, 148 GPa and 351 GPa, and 171 GPa, 96 GPa and 243 GPa for $B_{12}S$ and $B_{12}Se$, respectively, obtained using the GGA functional. As $B$, $G$ and $Y$ measure the volume deformation resistance, shear deformation resistance and stiffness of solids, $B_{12}S$ exhibits strong volume and shear deformation resistances as well as strong stiffness compared to $B_{12}Se$.

Next, we discuss the brittle/ductile behavior of the studied compounds. Brittleness/ductility is a decisive property for application purpose. The related parameter can be obtained based on the Poisson's ratio ($v$) and Pugh ratio ($G/B$). The $v$ and $G/B$ have an opposite tendency to have the values owing to the interdependence of the polycrystalline elastic parameters. The Poisson's ratio, used to measure the lateral structural deformation during the stretching or compression, is widely used to predict the ductility or brittleness of solids with a critical value of $v \sim 0.26$ (for brittle materials $v < 0.26$ and for ductile materials $v > 0.26$).[55–58] As predicted from the calculated values in Table 4, $B_{12}S$ should behave as a brittle material whereas for $B_{12}Se$ the Poisson's ratio has the values of 0.26 and 0.22 using the GGA and LDA functionals, respectively. Thus, we can define $B_{12}Se$ as a quasi-brittle material. Again, for the Pugh ratio ($G/B$), the critical value used to separate ductile and brittle behavior is 0.57 [for brittle materials $G/B > 0.57$ and for ductile materials $G/B < 0.57$].[59] Like Poisson's ratio, Pugh ratio also classifies $B_{12}S$ as a brittle solid. For $B_{12}Se$, $G/B$ has a value of 0.56 and 0.68 using the GGA and LDA functionals, respectively, which again directs us to define $B_{12}Se$ as a quasi-brittle material. It is encouraging to find that results from both the indicators are fully consistent with each other.

Furthermore, we have calculated the hardness parameters using three well-established models, namely, Chen's model[60], Miao's model[61] and Mazhnik's model[62] using USPEX[63]. The results are given in Table 4. As can be seen from Table 4, the hardness parameters are higher for $B_{12}S$ compared to those for $B_{12}Se$. Among the mechanical characterizing parameters, $C_{44}$ is assumed to be the best hardness predictor parameter.[64] The value of $C_{44}$ for $B_{12}S$ is 119 GPa using GGA, 121 GPa using LDA. These values much higher than those of $B_{12}Se$ (86 GPa using GGA, 99 GPa using LDA). Therefore, higher values of hardness parameters are expected for $B_{12}S$ compared to $B_{12}Se$. The higher hardness values of $B_{12}S$ compared to $B_{12}Se$ can also be explained on the basis of the DOS as shown in Fig. 12. A red rectangle is drawn in the figure in which there is a large difference in the values of TDOS for $B_{12}S$ and $B_{12}Se$. As seen, a peak in the DOS very close to the Fermi energy ($E_F$) is decreasing to a certain value (2.30 states/eV) and then there is an upturn to exhibit another peak in the lower energy part for $B_{12}S$. On the other hand, the peak close to $E_F$ is reduced to almost zero (0.38 states /eV) and then the upturn exhibits another peak in the lower energy part for $B_{12}Se$. In this region, a strong hybridization between B-$2p$ and S/Se-$3p$ electronic orbitals are observed that leads to the formation of covalent bonds between B and S/Se atoms. The strength of this hybridization depends largely on the value of the density of states, more states involved in the hybridization, stronger is the covalent bond formed. Thus, the covalent bond formed between B and S in this region is much stronger than the covalent bond formed between B and Se atoms in this energy region. Therefore, a higher value of hardness is expected for $B_{12}S$ compared to $B_{12}Se$, which is reflected well in the values presented in Table 4.

Now, we can compare the obtained values of hardness for titled compounds with some other B-rich compounds ($ZrB_{12}$, $HfB_{12}$, $YB_{12}$, $LuB_{12}$, $WB_{12}$, $TiB_{12}$). Korozlu et al.[12] have reported the hardness values (using Gao's model[65]) of 40.1 GPa, 39.1 GPa, 36.5 GPa, 32.3 GPa for $ZrB_{12}$, $HfB_{12}$, $YB_{12}$ and $LuB_{12}$, respectively. However, Pan et al.[13] have reported the hardness values of

29.9 GPa and 43.2 GPa for $WB_{12}$ and $TiB_{12}$, respectively using Miao's model. The hardness values of above mentioned compounds are higher than those of our considered compounds. Some of them can be considered as super-hard materials with hardness greater than 40 GPa ($ZrB_{12}$ and $TiB_{12}$) and rest of them are considered as hard materials. We have also studied some B-rich compounds ($B_6X$; X = S, Se) with hardness values of 33.60 GPa and 35.57 GPa (using Chen's formula) and categorized those as hard materials. Mazhnik et al.[62] have also calculated the hardness of widely known hard materials such as $B_4C$ (32.6 GPa), $B_6O$ (35.5 GPa), β-SiC (34.8 GPa), $SiO_2$ (30.0 GPa), WC (33.5 GPa), $OsB_2$ (17.8 GPa), VC (26.5 GPa), $Re_2B$ (38.6 GPa), etc. It is seen that the calculated values of hardness for both the titled compounds using the aforementioned three models are significantly varied, ranging within 19-31 GPa and 11-24 GPa for $B_{12}S$ and $B_{12}Se$, respectively. Thus, by comparing the hardness values of the titled chalcogenides with those of the other mentioned compounds, we might classify $B_{12}S$ as a potential hard material. But the hardness of $B_{12}Se$ is much lower than all the other compounds and it is less likely to be a hard material. However, the results of hardness presented herein demand experimental verification.

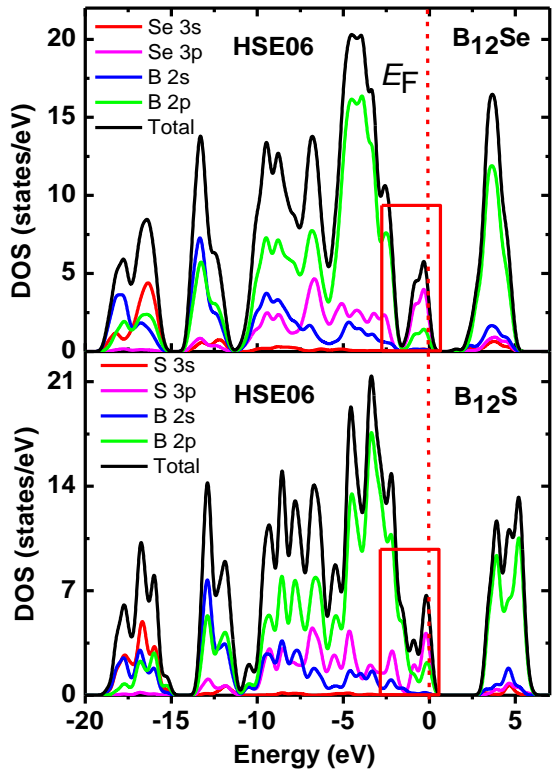

**Fig. 12.** Total and partial DOS for $B_{12}S$ and $B_{12}Se$ compounds as a function of electron energy. A red rectangle is drawn in the figure showing the difference in the values of DOS for both the compounds.

For hard materials, fracture toughness ($K_{IC}$) also carries significant importance like hardness for practical applications of solids.[66] Fracture may occur within a solid under extreme conditions of external stress. In such cases, prior information regarding fracture toughness is crucially important because it provides the required guideline for use and gives a measure of the resistance of solids to stop the propagation of the induced fracture inside. High value of fracture toughness is expected for hard materials for use in industrial purposes and evaluation of the $K_{IC}$ receives significant scientific concerns.[67–69] The obtained values of the hardness parameters inspire us to study the $K_{IC}$ of $B_{12}X$ (X = S, Se) using the formula, $K_{IC} = V_0^{1/6} \cdot G \cdot (B/G)^{1/2}$.[62] The calculated values of $K_{IC}$ are 4.27 MPam$^{0.5}$ and 3.13 MPam$^{0.5}$ for $B_{12}S$ and $B_{12}Se$, respectively using the GGA while those are 4.55 MPam$^{0.5}$ and 3.80 MPam$^{0.5}$ obtained using the LDA. These values are higher than those of $B_6S$ (2.070 MPam$^{0.5}$) and $B_6Se$ (2.072 MPam$^{0.5}$).[6] Moreover, Mazhnik et al.[62] have also reported the $K_{IC}$ of super hard diamond (6.3 MPam$^{0.5}$), WC (5.4 MPam$^{0.5}$) and c-BN (5.4 MPam$^{0.5}$) which are higher than those of $B_{12}S$ and $B_{12}Se$. Overall, the fracture hardness of boron-rich $B_{12}S$ and $B_{12}Se$ are reasonably high.

The hard materials are suitable for structural components in different sectors, but one of the great problems is their low level of machinability. A hard material with higher machinability is a great combination for engineering applications. Thus, information regarding machinability index (MI) calculated using $B/C_{44}$ ratio is crucial for predicting applications of hard materials.[6,70] The obtained values of MI are 1.57 (1.66) and 1.98 (1.85) for $B_{12}S$ and $B_{12}Se$ using GGA (LDA) functional. These values are higher than those of $B_6S$ and $B_6Se$. These values are also higher than those of some well-known machinable MAX phases[70] which predict that $B_{12}S$ and $B_{12}Se$ might be quite machinable.

The melting temperature, $T_m$, is estimated for both the studied chalcogenides using the following formula: $T_m = 412 + (8.2 \times C_{11})$.[71] The obtained values of $T_m$ are 4208 (4405) K and 3577 (3995) K for $B_{12}S$ and $B_{12}Se$ compounds using the GGA (LDA) functional, respectively. These values are higher than those of $B_6S$ and $B_6Se$.[6] The compound $B_{12}S$ have higher $T_m$ than $B_{12}Se$ and considering the computational uncertainty for using different functionals it can be stated that the $B_{12}S$ and $B_{12}Se$ compounds should be stable at least up to 4208 K and 3577 K, respectively, which are much higher value than the synthesis temperature of 2500 K.[8] According to the Born criterion,[72] the $T_m$ of a material is related to the shear modulus ($G$) as well as elastic stiffness constant $C_{44}$. Melting starts of a material when $G$ vanishes and the material becomes elastically unstable.[73] Higher values of $G$ and $C_{44}$ correspond to the higher value of $T_m$. Our calculated values of $T_m$ for both the compounds are consistent with the estimated $G$ and $C_{44}$ values as shown in Table 4.

3.5. Elastic Anisotropy

Elastic anisotropy should be taken into account for full description of mechanical properties of solids, specially for hexagonal symmetry where the atomic arrangement varies largely along the $a(b)$- and $c$-axes. Moreover, it is also closely associated with unusual phonon modes, dislocation dynamics, precipitation, plastic deformation in solids, and micro-scale crack creation.[74–76] Thus, information regarding elastic anisotropy of solids is very important for use in materials engineering and in crystal physics. Unfortunately, no unique method has been developed to measure the degree of anisotropy yet, and for this reason we have used different models which are widely used.

First of all, we started with the calculation of shear anisotropy factors which are defined for hexagonal solids like $B_{12}S$ and $B_{12}Se$ as follows: $A_1 = \frac{1/6(C_{11}+C_{12}+2C_{33}-4C_{13})}{C_{44}}$, $A_2 = \frac{2C_{44}}{C_{11}-C_{12}}$, $A_3 = A_1 \cdot A_2 = \frac{1/3(C_{11}+C_{12}+2C_{33}-4C_{13})}{C_{11}-C_{12}}$[77] for the {100}, {010} and {001} planes in between $\langle 011 \rangle$ and $\langle 010 \rangle$, $\langle 101 \rangle$ and $\langle 001 \rangle$, and $\langle 110 \rangle$ and $\langle 010 \rangle$ directions, respectively. A unit (1) value of $A_{\{100\}}$, $A_{\{010\}}$ and $A_{\{001\}}$ indicates the completely isotropic nature otherwise. Values deviating from unity imply anisotropy. Degree of deviation measures the level of anisotropy. The calculated shear anisotropy factors are listed in Table 5 which shows that $B_{12}S$ and $B_{12}Se$ are anisotropic solids.

The anisotropy in bulk modulus and compressibility are defined as: $B_a = a\frac{dP}{da} = \frac{\Lambda}{2+\alpha}$, $B_c = c\frac{dP}{dc} = \frac{B_a}{\alpha}$, where $\Lambda = 2(C_{11} + C_{12}) + 4C_{13}\alpha + C_{33}\alpha^2$ and $\alpha = \frac{(C_{11}+C_{12})-2C_{13}}{C_{33}-C_{13}}$[77] and $\frac{k_c}{k_a} =$

$C_{11} + C_{12} - 2C_{13}/(C_{33} - C_{13})$, where $B_a$ and $B_c$ are bulk moduli along $a$- and $c$-axes; and $k_c$ and $k_a$ are the compressibilities along the $a$- and $c$-axes. The listed values (Table 5) of $B_a$ and $B_c$ ($B_a = B_c$ for isotropic nature) and $k_c/k_a$ ($k_c/k_a = 1$ for isotropic nature) confirmed the anisotropic nature of $B_{12}S$ and $B_{12}Se$.

Chung and Buessem[78] have introduced the percentage anisotropy in compressibility and shear which are defined as: $A_B = \frac{B_V - B_R}{B_V + B_R}$ and $A_G = \frac{G_V - G_R}{G_V + G_R}$. For isotropic solids $A_B$ ($A_G$) = 0, otherwise anisotropic nature is indicated. It is evident from Table 5 that both $B_{12}S$ and $B_{12}Se$ are slightly anisotropic.

The anisotropic nature of $B_{12}S$ and $B_{12}Se$ is further revealed by calculating the universal anisotropy index, $A^U$ using the relation[79]: $A^U = 5\frac{G_V}{G_R} + \frac{B_V}{B_R} - 6 \geq 0$, where the $V$ and $R$ denote the upper limit (Voigt, $V$) and lower limit (Reuss, $R$) of bulk and shear moduli. The obtained values of $A^U$ also confirmed that both $B_{12}S$ and $B_{12}Se$ are mechanically anisotropic. From the listed values of anisotropy constants as presented in Table 5 it is seen that $B_{12}Se$ is more anisotropic compared to $B_{12}S$.

**Table 5:** Calculated values of different anisotropy indices: shear anisotropy factors ($A_{100}$, $A_{010}$ and $A_{001}$), anisotropy in bulk modulus ($B_a$, $B_c$), compressibility ($k_c/k_a$), percentage anisotropy ($A_B$ and $A_G$) and universal anisotropy index $A^U$ of boron-rich chalcogenide, $B_{12}X$ (X = S, Se) compounds.

| Phases | $A_{100}$ | $A_{010}$ | $A_{001}$ | $B_a$ | $B_c$ | $k_c/k_a$ | $A_B$ | $A_G$ | $A^U$ |
|---|---|---|---|---|---|---|---|---|---|
| $B_{12}S$ | 1.48 | 0.67 | 0.99 | 655 | 422 | 1.55 | 1.52 | 4.02 | 0.45 |
| $B_{12}Se$ | 1.90 | 0.73 | 1.39 | 614 | 368 | 1.67 | 2.11 | 19.52 | 2.46 |

In the second part of this section, we have assessed the degree of anisotropy graphically for Young's modulus, compressibility, shear modulus and Poisson's ratio and summarized those in

Table 6. The study is completed by plotting the Young's modulus, compressibility, shear modulus and Poisson's ratio in 2D and 3D using the ELATE code.[80] The obtained 2D and 3D plots are presented in Fig.13(a-d) for $B_{12}S$. Plots for $B_{12}Se$ are qualitatively similar and are not shown. Fig. 13(a) demonstrates the anisotropic nature of Young's modulus ($Y$). As can be seen from the 2D plot, $Y$ is isotropic in the $xy$-plane but anisotropic in $xz$- and $yz$-planes. The maximum as well as minimum values are observed in $yz$-plane at different angles from the vertical or horizontal axis. The ratio between maximum and minimum values of $Y$ is 1.69 and 3.22 for $B_{12}S$ and $B_{12}Se$, respectively, i.e., $B_{12}Se$ is significantly more anisotropic compared to $B_{12}S$ as far as Young's modulus is concerned.

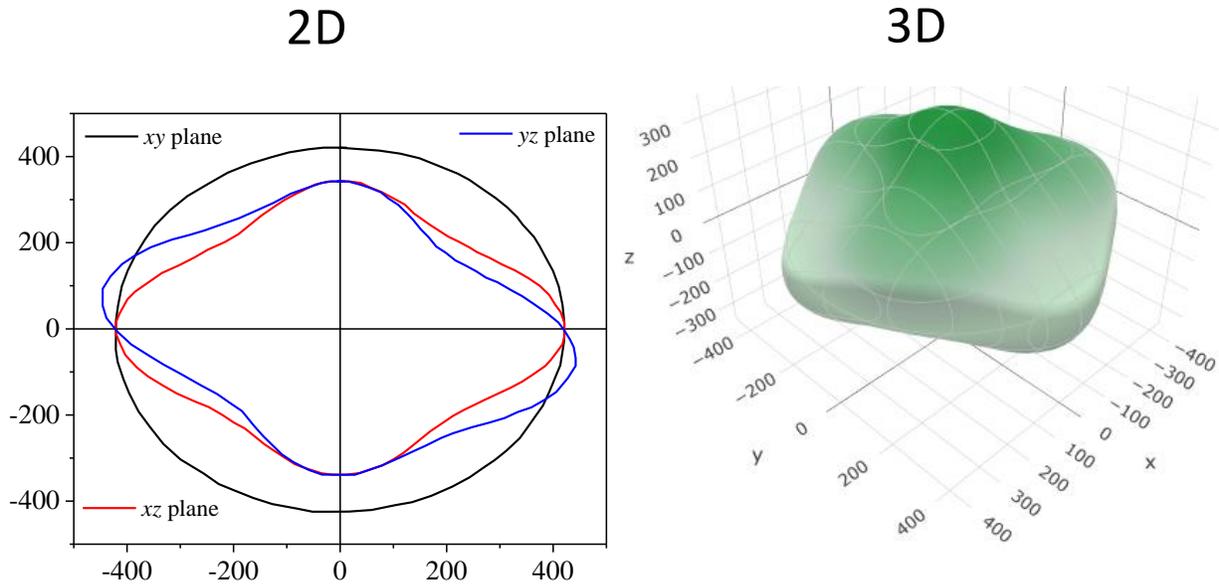

**Fig.13(a).** 2D view and 3D view of Young modulus ($Y$) for the chalcogenide $B_{12}S$ indicating the degree of anisotropy.

Fig. 13(b) illustrates the anisotropic nature of compressibility ($K$) which is isotropic in $xy$-plane like $Y$. Both $xz$- and $yz$-planes exhibit similar anisotropic nature as well as degree in the anisotropy. For both the planes, the maximum of $K$ lies on the vertical axis and minimum on the horizontal axis. The ratio between maximum and minimum values of $K$ is 1.55 and 1.66 for $B_{12}S$ and $B_{12}Se$, respectively, i.e., $B_{12}Se$ is slightly more anisotropic compared to $B_{12}S$ for compressibility.

**Table 6:** The minimum and the maximum values of the Young's modulus ($Y$), compressibility ($K$), shear modulus ($G$), and Poisson's ratio ($v$) of $B_{12}S$ and $B_{12}Se$ compounds.

| Phases | $Y_{min.}$ (GPa) | $Y_{max.}$ (GPa) | $A_Y$ | $K_{min}$ ($TPa^{-1}$) | $K_{max}$ ($TPa^{-1}$) | $A_K$ | $G_{min.}$ (GPa) | $G_{max.}$ (GPa) | $A_G$ | $v_{min.}$ | $v_{max.}$ |
|---|---|---|---|---|---|---|---|---|---|---|---|
| $B_{12}S$ | 265.16 | 448.71 | 1.69 | 1.52 | 2.37 | 1.55 | 105.76 | 190.17 | 1.79 | 0.006 | 0.340 |
| $B_{12}Se$ | 119.67 | 385.62 | 3.22 | 1.63 | 2.72 | 1.66 | 43.30 | 160.36 | 3.70 | -0.23 | 0.85 |

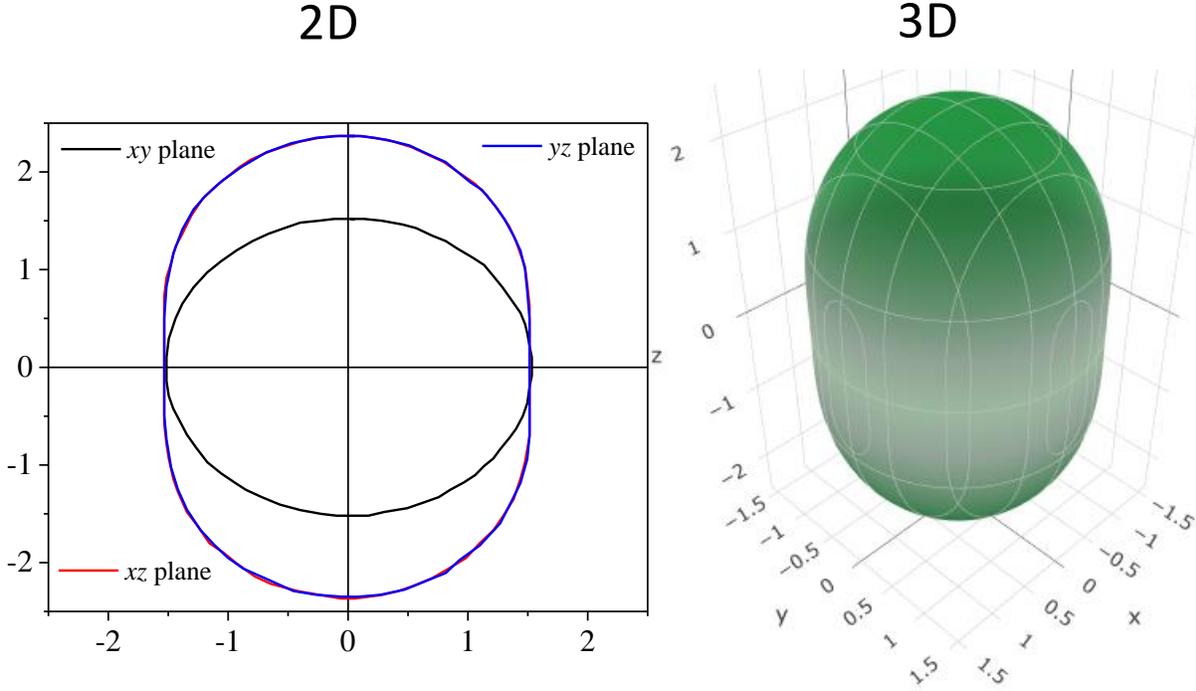

**Fig.13(b).** 2D view and 3D view of compressibility ($K$) for the chalcogenide $B_{12}S$ indicating the degree of anisotropy.

Fig.13(c) displays a complex anisotropy feature in shear modulus ($G$); different from $Y$ and $K$. The $G$ is not isotropic in any plane but exhibits two values for each point of the planes. The blue color indicates the maximum values while the red color indicates the minimum values. The maximum of $G$ is obtained on the horizontal axis of both $xy$- and $xz$-planes for both the compounds. For $yz$-plane, the maximum of $G$ lies at some angle above the positive horizontal axis as well at some angle below the negative horizontal axis for $B_{12}S$. The maximum and minimum values of $G$ lay almost midway of both horizontal and vertical axes in $yz$-plane for $B_{12}Se$. The ratio between maximum and minimum values of $G$ is 1.79 and 3.70 for $B_{12}S$ and $B_{12}Se$, respectively, i.e., $B_{12}Se$ is significantly more anisotropic compared to $B_{12}S$ for $G$. The

Poisson's ratio reveals similar type of anisotropy as $G$, two curves for each plane but much more complex than that of $G$. For this case, the maximum and minimum values are also revealed by the outer and inner curves at each angle as shown in Fig.13(d).

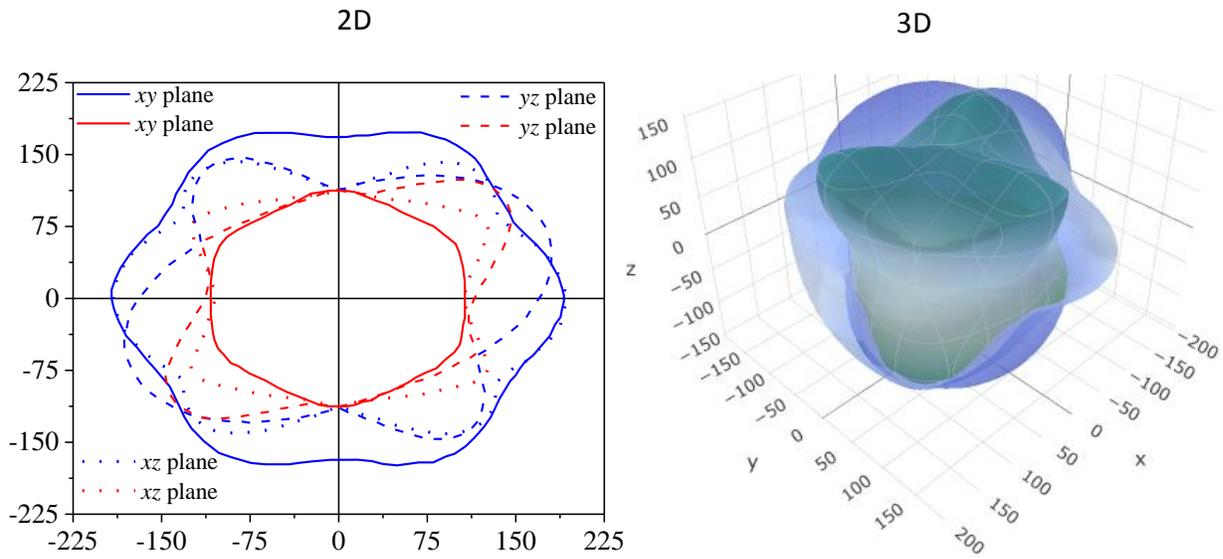

Fig. 13(c). 2D view and 3D view of shear modulus ($G$) for the chalcogenide $B_{12}S$ indicating the degree of anisotropy.

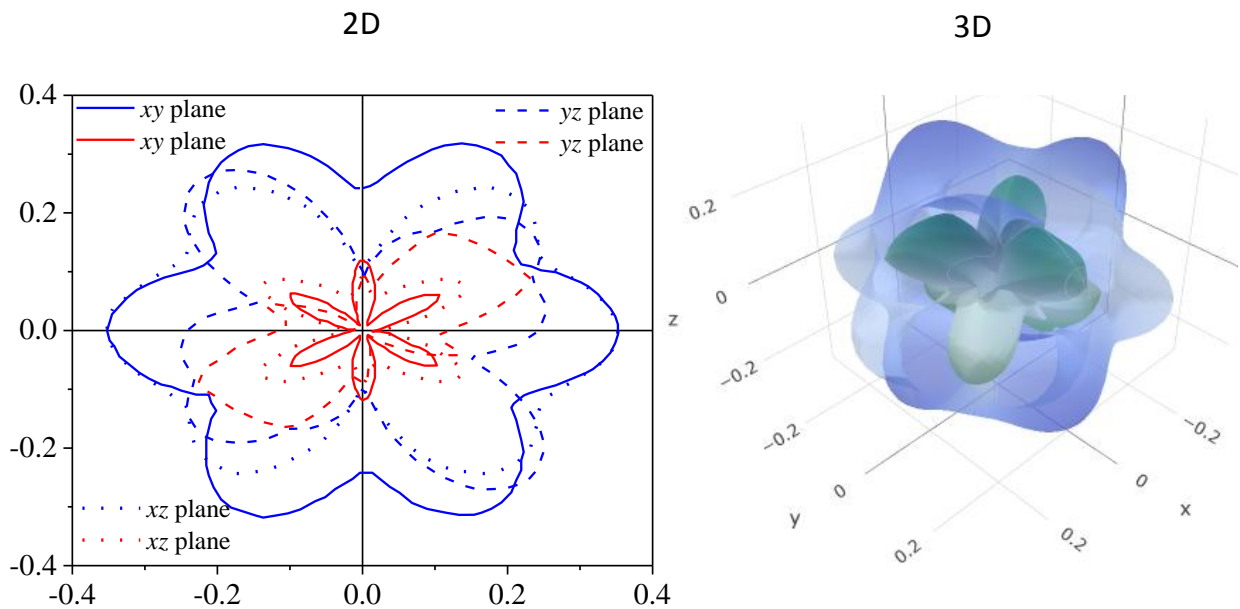

Fig.13(d). 2D view and 3D view of Poisson's ratio ($v$) for the chalcogenide $B_{12}S$ indicating the degree of anisotropy.

## 4. Conclusions

Density functional theory, incorporating different exchange-correlation functionals, is employed to characterize the structural, electronic band structure, optical (dielectric function, refractive index, absorption coefficient, reflectivity and loss function) properties of boron sub-sulfide ($B_{12}S$) and boron sub-selenide ($B_{12}Se$) compounds. Details of mechanical properties such as different elastic moduli, Pugh's ratio, Poisson's ratio, various anisotropy parameters, fracture toughness, melting point and machinability index have been estimated and discussed for the first time. The band gap is indirect in nature and its value is exchange-correlation functional sensitive. Large anisotropy in optical properties in the visible light range to mid-UV is observed. The absorption coefficient for electric field polarization along [100] of $B_{12}Se$ at visible region is found to be $3.25 \times 10^4$ cm$^{-1}$ that makes it competitive with renowned absorbance materials used in solar cell devices. The static value of refractive index is comparable with some commercialized compounds such as corundum, silica glass, quartz and soda lime glass. Lowest reflectivity spectra starts with value of 0.0742 (7.42 %) for $B_{12}S$ and 0.0815 (8.15 %) for $B_{12}Se$ compound for [001] polarization using the HSE06 functional. Such a pronounced absorption coefficient and very low value of reflectivity spectra along with bulk optical anisotropy at visible light region are crucially important to design many optical devices such as LCD screens, screens for 3D movies, polarizers, wave plates etc. Moreover, tunable band gaps of 2.27 and 1.30 eV for $B_{12}S$ and $B_{12}Se$, respectively, using the HSE06 functional in the visible light region also augurs well for solar cell device application although a thorough efficiency calculation is required. From the point of view of mechanical properties, the studied compounds have strongly anisotropic characteristics with weakly brittle nature confirmed from the shear to bulk modulus ratio and Poisson's ratio. The calculated values of fracture toughness are 3.13 and 3.80 MPam$^{0.5}$, at least, for $B_{12}S$ and $B_{12}Se$, respectively, indicating high degrees of resistance to limit the propagation of the induced fracture in these two solids. The values of mechanical hardness using Chen's model, Miao's model and Mazhnik's model are in the range of (19-31 GPa) and (11-24 GPa) for $B_{12}S$ and $B_{12}Se$, respectively. The origin of these high hardness values are successfully explained by considering the hybridization dependent density of state values at the valence band near the Fermi level. Almost all the results disclosed in this work are novel in nature and should act as useful references for further investigations. We believe that our predicted results will inspire both experimentalists as well as theorists for further study of the physical properties of $B_{12}S$ and $B_{12}Se$ in near future.

**Acknowledgements**

Authors are grateful to the Department of Physics, Chittagong University of Engineering & Technology (CUET), Chattogram-4349, Bangladesh, for providing the computing facilities for this work.

**Data availability**

The datasets generated during the current study are available from the corresponding authors on a reasonable request.

**Conflict of Interest**

The authors declare that they have no known competing financial interests or personal relationships that could have appeared to influence the work reported in this paper.